\documentclass [english,aps,prb,twocolumn,amsmath,amssymb,showpacs,showkeys,nofootinbib]{revtex4}
\usepackage[T1]{fontenc}
\usepackage[latin1]{inputenc}
\usepackage{graphicx}
\usepackage{gensymb}
\makeatletter

\usepackage{epsfig}
\usepackage{prettyref}
\usepackage{babel}
\usepackage{subfigure}
\makeatother
\begin{document}
\title{Unusual superparamagnetic behavior of Co$_{3}$O$_{4}$ nanoparticles}
\author{Vijay Bisht}
\email{vijayb@iitk.ac.in}
\author{K.P.Rajeev}
\email{kpraj@iitk.ac.in}
\affiliation {Department of Physics,
Indian Institute of Technology Kanpur 208016, India}

\begin{abstract}
We report  detailed studies on  magnetic properties of
Co$_{3}$O$_{4}$ nanoparticles of average size 12.5 nm. Temperature
and field dependence of magnetization, wait time dependence of
magnetic relaxation (aging), memory effects and temperature
dependence of specific heat have been investigated  to understand
the magnetic behavior of these particles. We find that the
particles show some features characteristic of nanoparticle
magnetism such as bifurcation of field cooled (FC) and zero field
cooled (ZFC) susceptibilities and a slow relaxation of
magnetization. However, strangely, the temperature at which ZFC
magnetization peaks coincides with the bifurcation temperature and
does not shift on application of magnetic fields up to 1~kOe,
unlike most other nanoparticle systems. Aging  effects in these
particles are negligible in both FC and ZFC protocol and memory
effects are present only in FC protocol. Our results show that
Co$_{3}$O$_{4}$ nanoparticles constitute a unique system where
superparamagnetic blocking starts above the Néel temperature, in
the paramagnetic state.

\end{abstract}

 \maketitle

\section{INTRODUCTION}

In the past two decades, magnetic nanoparticles have attracted
much attention due to (1) their potential uses in various areas
such as data storage and biomedicine and (2) the challenge to
understand the physics underlying the various exotic phenomena
exhibited by them.\cite{Dormann,Steen} Most of such work has
focussed on ferromagnetic and ferrimagnetic nanoparticles because
of their high magnetic moments that make them industrially
valuable. However, there have been a few studies on
antiferromagnetic nanoparticles and interestingly, their magnetic
behavior is found to be more complex and intriguing.

 Below a critical size, particles of a ferromagnetic or ferrimagnetic
material consist of a single domain and each particle carries a
magnetic moment, which can reverse its direction due to thermal
agitation. In these particles, the magnetic dynamics is thermally
activated leading to paramagnetic behavior above a particular
temperature, called blocking temperature. Below this temperature,
the moments appear blocked in a particular direction, on the time
scale of the measurement. This phenomenon is known as
superparamagnetism as instead of atomic spins, particle moments
("superspins") are involved. Thus ferromagnetic nanoparticles are
expected to show superparamagnetism, though, interparticle
interactions can complicate matters leading to spin glass like
behavior.\cite{Bisht,Batlle,Sasaki,Sun,Malay,Sahoo}
Antiferromagnetic nanoparticles can also develop a net moment due
to uncompensated spins and can exhibit superparamagnetism as
proposed by Néel.\cite{Neel,Brown} However, surface effects can
also be important in the determination of their magnetic behavior
and spin glass like behavior can arise due to freezing of spins at
the surface in addition to interparticle
interactions.\cite{Tiwari}

Magnetic nanoparticles show some characteristic features which are
common to both spin glasses and superparamagnets. These include
irreversibility in the field cooled (FC) and zero field cooled
(ZFC) magnetizations, a peak in ZFC magnetization, slow magnetic
relaxation and hysteresis at low temperature. However, some
important features that distinguish these two are: (1) FC
magnetization goes on increasing as the temperature is decreased
in superparamagnets while it tends to saturate below the freezing
temperature in particles showing spin glass behavior.\cite{Sasaki}
(2) In systems showing spin glass like behavior, wait time
dependence of relaxation (aging) and memory effects are present in
both FC and ZFC magnetizations. In contrast, in superparamagnetic
particles, these effects are present in FC magnetization
only.\cite{Bisht,Batlle,Suzuki,Sasaki,Sun,Malay,Chakraverty,
Tsoi,RZheng,Martinez,Sahoo} (3) The field dependence of
temperature at which the ZFC magnetization peaks ($T_P$) is known
to behave differently in the two
cases.\cite{Tiwari,Almeida,Bitoh,RKZheng}

There have been studies on various antiferromagnetic nanoparticles
in both bare and coated forms in the past several years and there
is a lot of variation in the magnetic behavior of different
materials. NiO nanoparticles have been reported to show spin glass
like behavior; \cite{Bisht,Tiwari} CuO nanoparticles show an
anomalous magnetic behavior that cannot be described as
superparamagnetic or spin glass like; \cite{Vijay} Ferritin shows
superparamagnetic behavior; \cite{Kilcoyne,Sasaki} etc.
Co$_{3}$O$_{4}$ is an antiferromagnetic  material and in the bulk
form, its  Néel temperature has been reported   to lie between
30~K and 40~K.\cite{Resnick}There have been some reports on
hysteresis, time dependence of magnetization, exchange bias and
finite size effects in bare, coated and dispersed Co$_{3}$O$_{4}$
nanoparticles and various conflicting claims have been made in
support of spin glass like and superparamagnetic behavior in these
particles. \cite{Yuko,Takada,Yuko1,Resnick,Makhlouf,Lin,Michi,
Mousavand,Shandong,Dutta}  It will be, therefore, worthwhile to
investigate their magnetic behavior carefully. In the present
work, we  present temperature and field dependence of
magnetization, aging and memory effects and specific heat
measurements on Co$_{3}$O$_{4}$ nanoparticles.

\section{EXPERIMENTAL DETAILS}

Co$_{3}$O$_{4}$ nanoparticles are prepared by a sol gel method.
Aqueous solution of sodium hydroxide is mixed with that of cobalt
nitrate till the pH of the solution becomes 12. At this stage
cobalt hydroxide  separates from the solution forming a gel that
is centrifuged to obtain a precipitate which is washed with water
and ethanol several times and  dried to obtain a precursor. This
precursor is heated at 250\degree C for 3 hours to obtain
Co$_{3}$O$_{4}$ nanoparticles. The sample is characterized by
X-ray diffraction (XRD) using a Seifert diffractometer with
Cu~K$\alpha$ radiation.  The average particle size as estimated
from the broadening of XRD peaks using the Scherrer formula comes
out to be 12.5~nm. All the magnetic measurements are done with a
SQUID magnetometer (Quantum Design). Specific heat measurements
are done with a PPMS(Quantum Design).

\section{RESULTS AND DISCUSSION}

\subsection{Temperature and field dependence of magnetization}

\begin{figure}[t]
 \begin{centering}
\includegraphics[width=1\columnwidth]{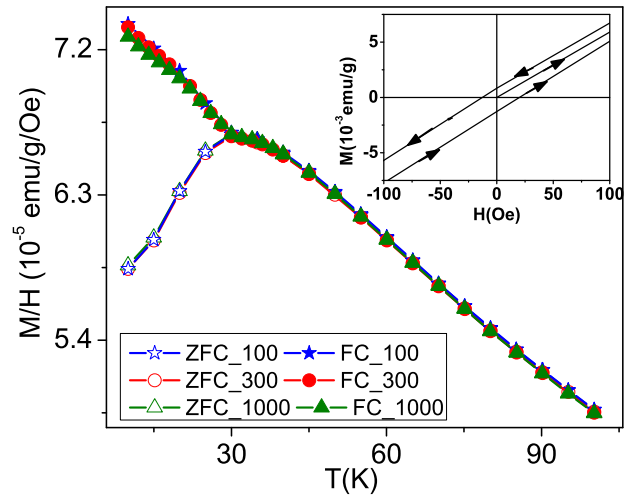}
\par\end{centering} \caption{(Color online) Temperature dependence
of FC and ZFC susceptibilities  at 100~Oe, 300~Oe and 1000~Oe.
Inset shows the hysteresis at 5~K.}
 \label{fig:FCZFC}
\end{figure}

The temperature dependence of magnetization was done under field
cooled (FC) and zero field cooled (ZFC) protocols at fields
100~Oe, 300~Oe and 1000~Oe. See Figure \ref{fig:FCZFC}. We note
that the FC magnetization in this case is increasing with decrease
in temperature down to the lowest temperature of measurement
without any sign of saturation, a feature characteristic of
superparamagnets.\cite{Sasaki} However, the temperature at which
ZFC magnetization peaks (31~K) i.e. $T_P$ is also the temperature
of bifurcation ($T_{bf}$) of FC and ZFC
magnetizations.\cite{footnote} Strangely, in this case, $T_P$ is
independent of the magnetic field applied and the dc
susceptibility data taken for different fields superpose. This is
in contrast to other nanoparticles where the peak temperature is
found to shift to lower temperatures on increasing the field even
at fields as low as a few hundred
Oeresteds.\cite{Bitoh,RKZheng,Almeida,Tiwari,Kachkachi,Sappey}
Another distinct feature in Figure \ref{fig:FCZFC} is a sudden
change in the slope of FC magnetization curve at the peak
temperature. This will be discussed later in subsection D.

 We have done hysteresis measurements at 5~K and a magnified view
 is shown in the inset of Figure \ref{fig:FCZFC}. The virgin curve
 in this case is a straight line in contrast to spin glasses where
 it is usually S-shaped.\cite{Mydosh}

\begin{figure}[b]
\begin{centering}
\includegraphics[width=1\columnwidth]{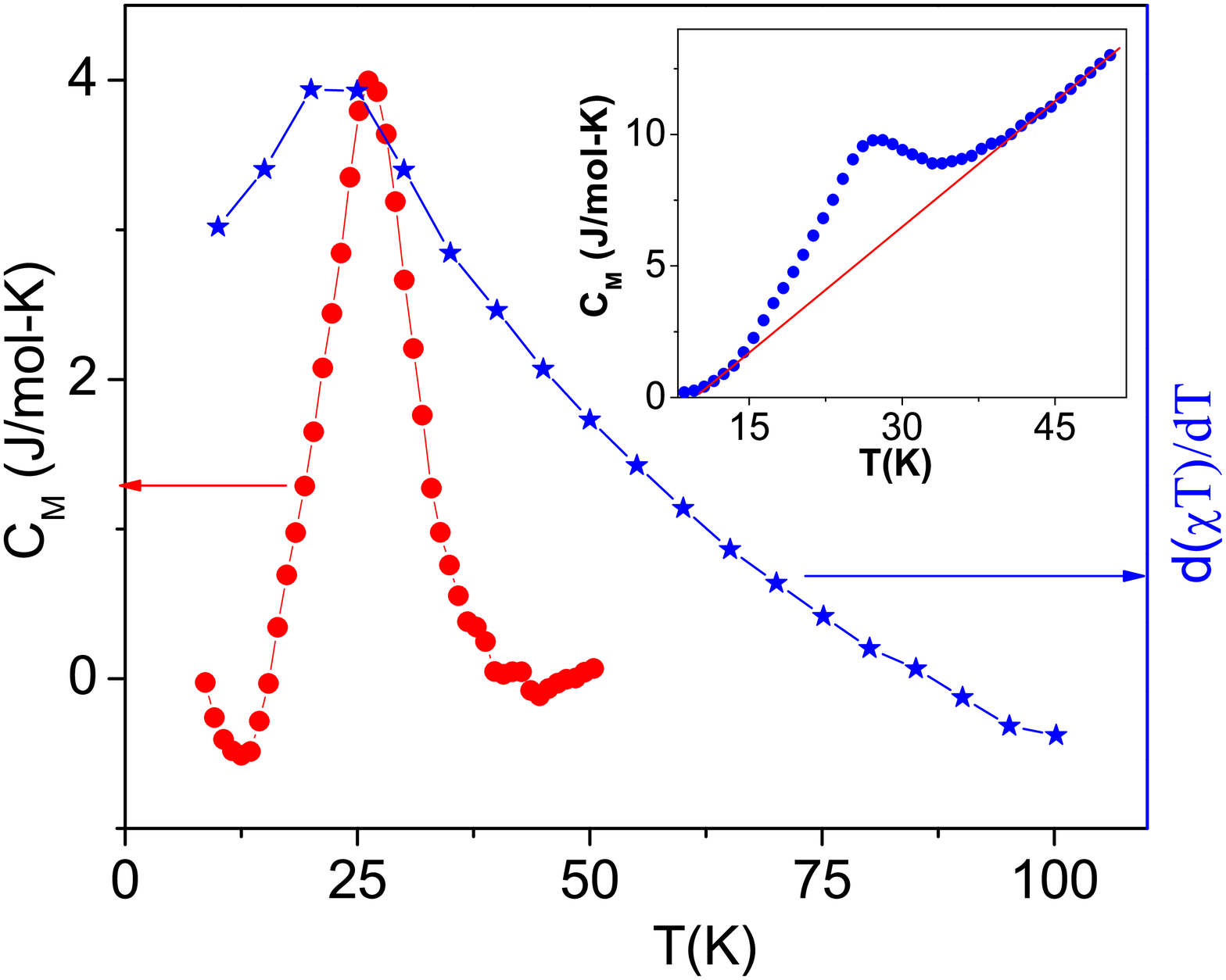}
\par\end{centering}
\caption{(Color online) Magnetic contribution to specific heat
estimated by subtracting the linear component at lower
temperatures and plot of  $d(\chi T)/dT$ as a function of
temperature. Inset shows the plot of specific heat as a function
of temperature. The line (red) shows the linear part of specific
heat which is subtracted from total specific heat to get the
magnetic contribution.} \label{fig:specheat}
\end{figure}

\subsection{Néel temperature}

When the particle size of an antiferromagnetic material is
decreased to nanoscale, its Néel temperature ($T_N$) is known to
decrease significantly due to  finite size
effects.\cite{Zheng,Lin} We carried out specific heat measurements
on pelletized  nanoparticles to make an estimate of $T_N$ and
these data are presented in the inset of
Figure~\ref{fig:specheat}. It can be observed that the specific
heat decreases linearly with decrease in temperature down to 35~K
below which it increases with a peak at 28~K. We have calculated
the magnetic specific heat  by subtracting a linear contribution
 from  the total specific heat and this data has
been shown in the main panel of Figure~\ref{fig:specheat}. It can
be seen that a magnetic contribution is noticeable between 15~K
and 40~K with a peak at 26~K, which should correspond to the Néel
temperature.

 In some works on antiferromagnetic nanoparticles,
$T_N$ has been identified as the temperature of the peak of $ d
(\chi T)/dT$ vs $T$ curve.\cite{Dutta,Punnoose} This is in
accordance with Fisher's equation relating specific heat (C) and
magnetic susceptibility ($\chi $):\cite{Fisher}
\begin{equation} \label{Fisher relation} C \propto  d
(\chi T)/dT \end{equation} This relation has been verified
experimentally for some bulk materials.\cite{Bragg} It will be
interesting to check this relation for the present nanoparticle
sample. In Figure \ref{fig:specheat}, we have also shown the plot
of $d(\chi T)/dT$ as a function of $T$, which has a peak somewhere
in between 20~K to 25~K, giving an estimate of the Néel
temperature. Thus the value of $T_N$ extracted from the specific
heat is somewhat greater than the value obtained from the Fisher
relation.

 \subsection{Aging and memory effects}

Nanoparticles that show spin glass like behavior are expected to
show aging and memory effects in both FC and ZFC protocols while
those showing superparamagnetic behavior  show these effects only
in FC protocol. Further the effects in FC protocol are weaker in
superparamagnetic particles than in those showing spin glass like
behavior.\cite{Bisht,Batlle,Sasaki} These experiments can thus
give valuable information about the nature of magnetic behavior of
a system.

\begin{figure}[t]
\begin{centering}
\includegraphics[width=1\columnwidth]{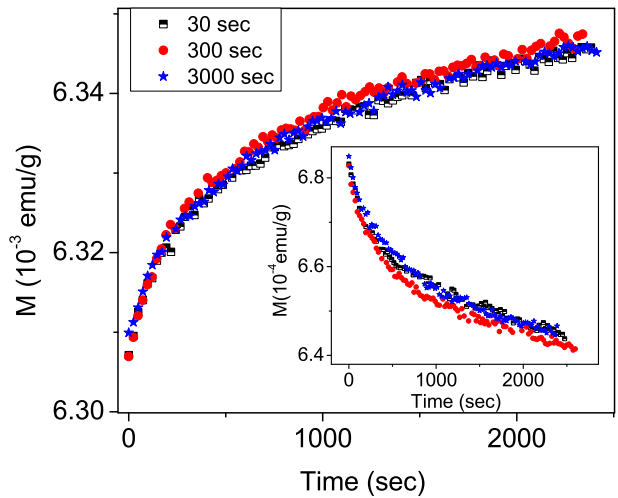}
\par\end{centering}
\caption{(Color online) Aging experiments in ZFC protocol at 20~K
with waiting times 30~s, 300~sec and 3000~sec. Inset shows the
corresponding experiments in FC protocol. } \label{fig:aging}
\end{figure}

For doing aging experiments in FC protocol, the sample is cooled
to a particular temperature in a field of 100~Oe, the field is
switched off after waiting for a specified period and
magnetization is recorded as a function of time.  In the
corresponding ZFC case, the sample is cooled to the temperature of
interest in zero field and the field is switched on after a
certain wait time and subsequently magnetization is recorded as a
function of time. We show the results of aging experiments at 20~K
in Figure~\ref{fig:aging}. It can be seen that aging effects are
negligible in both FC and ZFC measurements.

\begin{figure}[b]
\begin{centering}
\includegraphics[width=1\columnwidth]{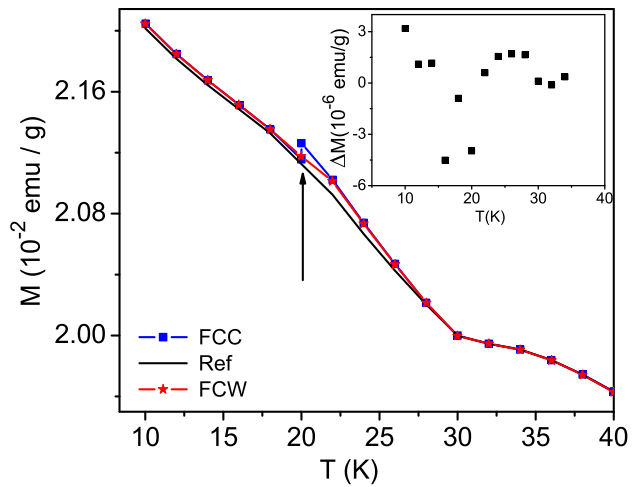}
\par\end{centering}
\caption{(Color online) Memory experiments in FC protocol with a
stop of one hour duration at 20~K at a field of 300~Oe. The field
is switched off during the stop (shown as arrow). Inset shows the
corresponding data for ZFC memory experiment. The difference in
magnetization with a stop of one hour at 20K in the cooling
process and the reference data is plotted as a function of
temperature. } \label{fig:FCmemory}
\end{figure}

 For carrying out FC memory experiments, the system is cooled in
 the presence of a magnetic field to 20~K
and a stop of one hour is taken at this temperature. Magnetic
field is switched off for the duration of the stop and is turned
on before further cooling the sample to 10~K. The magnetization is
measured while cooling and then during subsequent heating. These
data have been taken at  300~Oe field. See
Figure~\ref{fig:FCmemory}. It can be observed that there is some
indication of  memory as the heating curve meets the cooling curve
just after the temperature at which the stop was taken.  We have
also done memory experiments in ZFC protocol with a stop of one
hour at 20~K and did not find any indications of memory. See the
inset of Figure~\ref{fig:FCmemory}. In this case, $\Delta M $ is
less than $0.05\%$ as opposed to canonical spin glasses or
nanoparticle systems that show spin glass like behavior where
differences of order $1\%$ or more are
observed.\cite{Bisht,Matheiu}


\subsection{DISCUSSION}

Co$_{3}$O$_{4}$ nanoparticles show some features characteristic of
nanoparticle magnetism. Some of these are due to finite size
effects viz. a net magnetic moment due to uncompensated surface
spins and a decrease in Néel temperature. Some are manifestations
of non equilibrium in magnetic nanoparticles and are common to
both superparamagnetic and spin glass like behaviors. These
include irreversibility in FC and ZFC magnetization and a slow
magnetic relaxation at low temperature. However there are several
unusual features observed in this system, that deserve a second
look.

\subsubsection{Behavior above $T_P$}

In ferromagnetic and ferrimagnetic nanoparticles, the ZFC
magnetization generally shows a peak at a particular temperature,
above which the behavior is superparamagnetic i.e. magnetization
curves taken at various temperatures above $T_P$ superpose when
plotted against $H/T$.\cite{Bean}  Makhlouf et al. have shown that
magnetization vs $ H/(T$+$\theta$) curves superpose at high
temperatures in Co$_{3}$O$_{4}$ nanoparticles ($\theta = 85~K$), a
feature characteristic of antiferromagnets.\cite{Makhlouf}  Our
data (50~K-300~K) also fits well to Curie-Weiss law, $\chi \propto
1/(T+\theta)$,  with $\theta = 107~K$ and coefficient of
determination, $R^2 = 0.9992$.
 Further, $T_P$ is independent of applied field in
contrast to nanoparticles showing superparamagnetism. The
preceding discussion implies that the system is not
superparamagnetic above $T_P$ and the peak in ZFC magnetization is
a signature of a transition from paramagnetic to antiferromagnetic
state. It may be noted that $T_N$ as found by specific heat
measurements is 5~K less than $T_P$ and the Fisher relation gives
an even lower estimate of $T_N$. We note in passing that similar
observation of $T_N$ < $T_P$ has been reported
earlier.\cite{Bragg,Dutta}

At $T_P$, there also occurs a bifurcation between FC and ZFC
magnetization and this change looks abrupt as at this point a
slope change in the FC magnetization can be seen. Thus even before
the actual transition to an antiferromagnetic state, the particles
develop a magnetic moment and go to a blocked state. This may be
due to the short range antiferromagnetic correlations which are
known to persist above the Néel temperature.\cite{buschow}
 These short ranged correlations can give rise to a net magnetic
 moment when the correlation length becomes comparable to the particle
 size and thus the particles can get blocked above $T_N$.

\subsubsection{Behavior below  $T_P$}

 There have been some reports of spin glass like features in Co$_{3}$O$_{4}$
nanoparticles coated with organic surfactants and those dispersed
in  amorphous matrices at low temperature.
\cite{Resnick,Michi,Mousavand,Shandong} However, in the present
work and in other studies on bare nanoparticles, no such features
have been found. \cite{Makhlouf,Dutta} Absence of aging and memory
effects in ZFC protocol confirms that the behavior of
Co$_{3}$O$_{4}$ nanoparticles is not spin glass
like.\cite{Bisht,Sasaki,RZheng,Tsoi} Thus below $T_P$, observation
of a bifurcation in FC and ZFC magnetizations and slow relaxation
of magnetization seems to correspond to a blocked state as
observed in superparamagnetic particles. Presence of memory
effects in FC protocol also support this inference.

The blocking temperature in superparamagnets is known to be field
dependent, but in this case it is independent of field at least up
to fields of 1~kOe. In superparamagnetic particles, the particle
moment can have two directions along an axis determined by
anisotropy. In zero field, these two states are separated by an
energy barrier and are equally likely to be occupied. Application
of a magnetic field breaks the degeneracy of these two states,
 making it easier for the particle to go from the higher energy state (antiparallel
 to field) to the lower energy state (parallel to field).  At a
particular temperature, if the thermal energy available is
insufficient for crossing this barrier, the particles will be in a
blocked state. We propose that in the present case, particles
develop a magnetic moment at $T_P$ and immediately get blocked
with an energy barrier greater than $k_B T_P$. The applied
magnetic fields (up to 1 kOe in this case) are not sufficient to
decrease the barrier significantly so that the particles are not
able to get unblocked at this temperature. Hence the blocking
temperature in this case is independent of field.

Another unique feature observed in this system is that the ZFC
susceptibilities taken at different fields lie on the same curve
while there is a slight mismatch between corresponding FC
susceptibilities. This is due to the fact that in ZFC measurement
the system is cooled in a zero field and then a field is applied
at low temperature to measure the magnetization, and thus the
corresponding values of magnetization lie on the virgin curve of
hysteresis loop which is a straight line with zero intercept ($M/H
= $ constant). However in FC measurements, the system is cooled in
the presence of a field, and hence the magnetization values do not
lie in the virgin curve, leading to slightly different
susceptibilities at different fields.

\section{Conclusion}
  We have done a detailed study of the magnetic behavior of
Co$_{3}$O$_{4}$ nanoparticles. We find that their behavior is
unique among antiferromagnetic nanoparticles. There is a peak in
ZFC magnetization  and at this temperature ($T_P$), a sudden
bifurcation between FC and ZFC magnetization occurs. Strangely,
this temperature lies above the Néel temperature as estimated by
specific heat measurements and a sudden slope change in FC
magnetization is seen at this point. The behavior of
susceptibility above the peak temperature is paramagnetic rather
than superparamagnetic and Curie-Weiss law holds as in the case of
antiferromagnets. Further unlike other magnetic nanoparticles,
$T_P$ does not change with magnetic field. Aging and memory
effects are not observed in ZFC magnetization measurements which
show the absence of spin glass like behavior in this system.
However observation of memory in FC protocol supports
superparamagnetic blocking below $T_P$. This system undergoes a
transition from paramagnetic to blocked state even before the
transition to an antiferromagnetic state and we ascribe this
  to antiferromagnetic correlations present above Néel temperature.

 \begin{acknowledgments}
VB thanks the University Grants Commission of India for financial
support. Authors thank Prof. C .V .Tomy, IIT Bombay for specific
heat measurements.
\end{acknowledgments}


\begin{thebibliography}{40}
\bibitem{footnote} Some authors have observed a $T_{bf}$
which lies a few Kelvins above $T_P$and some others have found it
even below $T_P$. \cite{Dutta,Makhlouf}
\end{thebibliography}
\end{document}